\documentclass{ws-procs961x669}            % default, citations in superscript
\begin{document}
\title{Shadows of Kerr-like black holes in $4D$ Einstein--Gauss--Bonnet gravity and constraints from EHT observations}

\author{Sushant~G.~Ghosh}
\address{Centre for Theoretical Physics, Jamia Millia
		Islamia, New Delhi 110025, India\\
		Astrophysics Research Centre, School of Mathematics, Statistics and Computer Science,\\
		University of KwaZulu-Natal, Private Bag 54001, Durban 4000, South Africa\\
		E-mail: sghosh2@jmi.ac.in\\}

\author{Rahul Kumar Walia}
\address{Astrophysics Research Centre, School of Mathematics, Statistics and Computer Science,\\
		University of KwaZulu-Natal, Private Bag 54001, Durban 4000, South Africa\\
		E-mail: rahul.phy3@gmail.com}

\begin{abstract}
The M87* black hole shadow observation by the Event Horizon Telescope (EHT) has enabled us to test the modified gravity theories in the extreme-field regime and estimating the black hole parameters. Having this assertion, we	investigate the Kerr-like rotating black holes in $4D$ Einstein-Gauss-Bonnet (EGB) gravity and deduce their shadows. Considering the inclination angle $\theta_0=17^o$, we show that the EGB black hole shadows are smaller and more distorted than for the Kerr black holes. Modelling the M87* black hole as the EGB black hole, we predict the shadow angular size $35.7888\mu as\leq \theta_d\leq 39.6192\mu as$. The M87* black hole shadow angular size $\theta_d=42\pm 3\mu as$, within the 1$\sigma$ region, constrains the GB coupling parameter and the black hole spin parameter. Interestingly, the circularity deviation of the EGB black hole shadows is smaller than the bounded deduced for the M87* black hole.
\end{abstract}


\bodymatter

\section{Introduction}
The uniqueness of the Einstein tensor to describe gravity in the four-dimensional ($4D$) spacetime is dictated by the Lovelock theorem \cite{Lovelock:1972vz}. However, if one or more conditions in the Lovelock theorem are relaxed, then modifications to the Einstein-Hilbert action exist that lead to covariant, conserved, and second-order field equations and propagate only gravitational degrees of freedom and thus are free from the ghost instabilities. One such Lagrangian-based theory of gravity is Einstein-Gauss-Bonnet (EGB) gravity that exists in the $D\geq 5$ and is motivated by the heterotic string theory \cite{Lanczos:1938sf, Lovelock:1971yv}. EGB gravity supplements the Einstein-Hilbert action with quadratic corrections terms constructed from the curvature tensors invariants and reads as follows 
\begin{equation}
\mathcal{I}_{\text{EGB}}
=\frac{1}{16\pi G_D} \int\! d^{D\!}x \, \sqrt{-g}(\mathcal{L}_{\text{EH}}+ \alpha \, \mathcal{L}_{\text{GB}}),\label{action}
\end{equation}
with
\begin{equation}
\mathcal{L}_{\text{EH}}=R,\;\;\;\; \mathcal{L}_{\text{GB}}=R^{\mu\nu\rho\sigma} R_{\mu\nu\rho\sigma}- 4 R^{\mu\nu}R_{\mu\nu}+ R^2.
\end{equation}
Here, $\alpha$ is identified as the GB coupling constant and is related to the inverse string tension, making it positive-definite. The GB Lagrangian is a unique quadratic combination of the Riemann tensor that naturally emerges as a leading-order correction term in low-energy effective actions of heterotic string theory and $10D$ gauged supergravity. Boulware and Deser, in their seminal paper \cite{Boulware:1985wk}, obtained the first spherically symmetric and static black hole solution for the EGB theory, afterward several intriguing black hole solutions are obtained \cite{egb2,ghosh,egb}. It is worth mentioning that for $D<5$ GB Lagrangian $\mathcal{L}_{\text{GB}}$ turns into a total derivative, and thereby its contribution to the gravitational dynamics vanishes, rendering the theory indistinguishable from general relativity. However, in the presence of an additional non-minimally coupled scalar field dilaton with the canonical kinetic term, $\mathcal{L}_{\text{GB}}$ leads to the non-trivial gravitational dynamics \cite{Sotiriou:2013qea,Sotiriou:2014pfa,Doneva:2017bvd,Cunha:2019dwb} and the resulting theory is Horndeski or Galilean.

Ever since the formulation of the EGB gravity theory, its $4D$ regularization has been a topic of great interest. In this line of research, Tomozawa \cite{Tomozawa:2011gp} showed for the first time that the quantum corrections to gravity in a conformally flat metric in $4D$ appears as GB quadratic curvature forms, and the $4D$ black hole solution shows repulsive nature at $r\to 0$. In another attempt of regularization procedure, Cognolo \textit{et al.} \cite{Cognola:2013fva} used an ``entropic" dimensional reduction of EGB gravity to $D\to 4$ within the classical Lagrangian formulation. Lately, the interest in the $4D$ EGB gravity theory is re-surged due to the regularization approach proposed by Glavan and Lin \cite{Glavan:2019inb}; the GB coupling is re-scaled as $\alpha\to \alpha/(D-4)$ and the $4D$ EGB theory was obtained as the limit $D\to 4$ at the level of field equations. The aim for introducing this re-scaling is to generate a divergence that exactly cancels the vanishing contribution that the GB term makes to the field equations in $4D$. The extension to higher-order Lovelock gravity is presented in Refs.~\cite{Konoplya:2020qqh,Casalino:2020kbt}. Likewise, EGB theory is obtained in lower dimensions \cite{Ma:2020ufk,Hennigar:2020lsl}. Interestingly, the Glavan and Lin's static and spherically symmetric black hole solution \cite{Glavan:2019inb} matched with that obtained using the quantum correction by Tomozawa \cite{Tomozawa:2011gp}, and Cognolo et al. \cite{Cognola:2013fva}.

However, Glavan and Lin's claim \cite{Glavan:2019inb} that the resulting theory is of pure graviton was later proven to be spurious on several grounds. The covariant approach proposed in Ref.~\cite{Glavan:2019inb} is largely speculated to be valid only for specific higher-dimensional spacetimes with high degrees of symmetries, particularly maximally symmetric or spherically symmetric spacetimes. Recently, some studies have called into question the Glavan and Lin \cite{Glavan:2019inb} regularization procedure for the less-symmetric spacetimes and also reported several other inconsistencies in Refs.~\cite{Ai:2020peo,Hennigar:2020lsl,Shu:2020cjw,Gurses:2020ofy,Mahapatra:2020rds}. Following that, the GB contribution arising in higher dimensions could be renormalized in several ways to yield a non-trivial contribution also in $4D$, some even without re-scaling the GB coupling \cite{Lu:2020iav,Kobayashi:2020wqy,Hennigar:2020lsl,Casalino:2020kbt,Ma:2020ufk,Arrechea:2020evj,Aoki:2020lig}. Hennigar  {\it et al.} \cite{Hennigar:2020lsl} proposed another well defined $D \to 4$ limit of EGB gravity generalizing the previous work of Mann and Ross \cite{Mann:1992ar} in establishing the $D \to 2$ limit of general relativity and this regularization is applicable not only in $4D$ but also to $D<4$. These alternate regularization procedures of EGB theory, leading to a divergence-free $4D$ action, describe the scalar-tensor theory of gravity of the Horndeski type. These scalar-tensor models propagate the supplementary scalar mode in addition to the gravitational degree of freedom. Thus these alternate regularized theories are in line with Lovelock's theorem, as they introduce another dynamical field. Nevertheless, the spherically symmetric $4D$  black hole solution obtained in Ref.~\cite{Glavan:2019inb} remains valid for these regularised scalar-tensor theories \cite{Lu:2020iav,Hennigar:2020lsl,Casalino:2020kbt,Fernandes:2020nbq,Ma:2020ufk}. This means that $4D$ EGB gravity can be viewed as both a dimensionally reduced theory and as a gravitational theory that displays known quantum corrections. As a result, both the Glavan and Lin theory \cite {Glavan:2019inb} and scalar-tensor regularizations have received remarkable attention and more then 100 papers have been reported on $4D$ EGB gravity and its various solutions including their charged extension \cite{Fernandes:2020rpa,Singh:2020nwo}, rotating counterparts \cite{Wei:2020ght,Kumar:2020owy}, Vaidya-like radiating black holes \cite{Ghosh:2020vpc,Ghosh:2020syx}, regular black holes \cite{Kumar:2020xvu,Kumar:2020uyz}. The gravitational lensing of $4D$ EGB black holes have also been studied \cite{Islam:2020xmy,Heydari-Fard:2020sib,Jin:2020emq,Kumar:2020sag}.

The black hole shadow observations by the Event Horizon Telescope (EHT) Collaboration have unprecedentedly opened up an exciting arena to make a precision test of the gravitational theory in the strong and relativistic field regimes (in the vicinity of the unstable bound orbits around black holes) \cite{Akiyama:2019cqa,Akiyama:2019eap}. The EHT analysis suggested that, based on \textit{a priori} known estimates for the mass and distance from stellar dynamics, the M87* shadow size is consistent within $17\% $ for a $68\%$ confidence interval of the size predicted from the Kerr black hole general-relativistic-magneto-hydrodynamics (GRMHD) image \cite{Psaltis:2020lvx}. However, several other studies altogether have not entirely precluded the possibility of non-Kerr black holes \cite{Mizuno:2018lxz,Vincent:2020dij,Junior:2021atr}. Using the M87* shadow angular size, constraints are placed on the second post-Newtonian metric coefficients, which were inaccessible in the earlier weak-field tests at the Solar-scale \cite{Psaltis:2020lvx}. Therefore, it is both legitimate and timely to test the viability of the $4D$ EGB gravity theory using the M87* black hole shadow observations. This paper aims to present the detailed study of the rotating $4D$ EGB black hole shadow, parameter estimation of the black hole using the shadow observables, and constraining them using the M87* black hole shadow observed by the EHT. 

\section{Rotating $4D$ EGB black hole shadows}
\begin{figure}
	\begin{center}
		\includegraphics[scale=0.7]{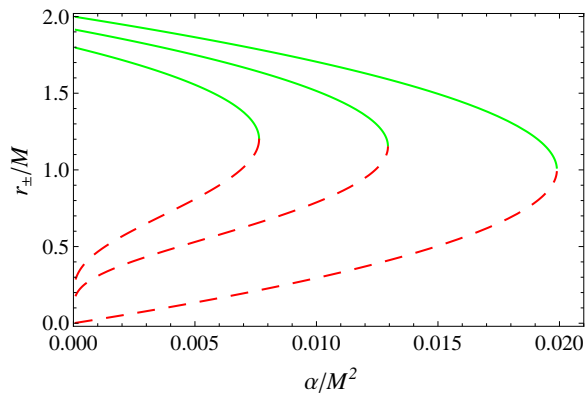}
	\caption{Event horizon (solid green) and Cauchy horizon (dashed red) radii variation with $\alpha$ for different values of $a=0, 0.40M, 0.60M$ (from outside to inside).}\label{horizons}
	\end{center}
\end{figure}

Finding an exact analytic and rotating axially symmetric black hole solution of the EGB gravity is a notorious task due to the non-linearity involved in the field equations. However, there exists the rotating solution generating mechanisms such as the Newman-Janis algorithm \cite{Newman:1965tw} and the gravitational-decoupling method \cite{Contreras:2021yxe}, which have been widely used to construct rotating black hole solutions from their non-rotating counterparts. The Azreg-A\"inou's non-complexification procedure\cite{Azreg-Ainou:2014pra,Azreg-Ainou:2014aqa} for the modified Newman-Janis algorithm generates a unique imperfect fluid rotating solution from the seed spherically symmetric static solution. It has been applied to generate rotating solutions in several modified gravity theories \cite{Johannsen:2011dh,Bambi:2013ufa, Ghosh:2014pba, Moffat:2014aja,Kumar:2017qws,Kumar:2020hgm,Kumar:2020owy}.
The rotating $4D$ EGB black hole metric, in Boyer-Lindquist coordinates, reads \cite{Kumar:2020owy,Wei:2020ght}
\begin{eqnarray}
ds^2 &=&  -\frac{\Delta}{\Sigma}\left(dt - a \sin^2 \theta\, d\phi\right)^2 + \frac{ \Sigma}{\Delta }  \, dr^2 + \Sigma \, d \theta^2 +  \, \frac{\sin ^2 \theta}{\Sigma}  \left((r^2+a^2)\,d\phi-a\,dt   \right)^2
\label{rotbhr}
\end{eqnarray}
with 
\begin{equation}
\Delta=r^2+a^2+\frac{r^4}{32\pi\alpha }\left[1- \sqrt{ 1+ \frac{128 \pi\alpha  M}{r^3} }\right] , \quad \Sigma=r^2+a^2\cos^2\theta.
\end{equation} 
Thus rotating $4D$ EGB black holes are characterized by three parameters, mass ($M$), spin ($a$), and GB coupling parameter ($\alpha$), which also gives potential deviation from the Kerr solution. In the limit $\alpha\to 0$ or large $r$, the metric Eq.~(\ref{rotbhr}) smoothly recovers the Kerr black hole \cite{Kerr:1963ud}. Because the static black hole solution of Ref.~\cite{Glavan:2019inb} is identical to that of regularized scalar-tensor theories and other quantum-corrected theories of gravity  \cite{Tomozawa:2011gp,Glavan:2019inb,Cai:2009ua,Cognola:2013fva,Casalino:2020kbt,Kehagias:2009is,Konoplya:2020qqh,Hennigar:2020lsl}, the rotating black hole metric (\ref{rotbhr}) also corresponds to these theories. The rotating black hole admits up to two distinct horizons whose radii $r_- \leq r_+$ can be identified as real positive roots of the $\Delta=0$. The variation of both horizon radii with GB coupling is shown in Fig.~\ref{horizons}, it is evident that the event horizon radius decreases and Cauchy horizon radius increase with $\alpha$. For a given value of spin $a$, there exists a extremal value of GB coupling $\alpha=\alpha_E$ for which degenerate horizons $r_-=r_+$ exists, such that for $\alpha>\alpha_E$ horizons disappear and the central singularity becomes globally naked. Similarly, for a given value of $\alpha$, one can find the extremal value of spin $a=a_E$ which leads to degenerate horizons $r_-=r_+$. In this paper, we will only consider the black hole case viz., $\alpha\leq \alpha_E$.

The optical appearance of the black hole in the presence of a bright background or the illuminated matter accretion flow is known as the shadow. The light from the source gets strongly lensed by the black hole in the vicinity of the horizon and receive by a faraway observer on the opposite side of the source. Synge \cite{Synge:1966}, and Luminet \cite{Luminet:1979} led the study of black hole shadow and calculated the capturing angle of the Schwarzschild black hole. For the first time, the shadow of the Kerr black hole was reported by Bardeen \cite{Bardeen} in his pioneering work in 1973. The rotating EGB metric (\ref{rotbhr}) belongs to Petrov type-D spacetimes, and thus the geodesics equations are completely integrable. The metric (\ref{rotbhr}) carries two Killing vectors $\partial_t$ and $\partial_{\phi}$, associated with the time translational and rotational invariance of the spacetime geometry. The components of photon four-momentum $p^{\mu}$ projected along these Killing vectors are constant of motion, which in this case can be identified as the energy $E$ and axial angular momentum magnitude $L_z$. We follow the Hamilton-Jacobi formalism to determine the null geodesics equations of motion around the rotating black hole, which read as follows \cite{Kumar:2020owy,Chandrasekhar:1992}
\begin{eqnarray}
\Sigma \frac{dt}{d\tau}&=&\frac{r^2+a^2}{\Delta}\left({ E}(r^2+a^2)-a{ L_z}\right)  -a(a{ E}\sin^2\theta-{{L}_z}) ,\label{tuch}\\
\Sigma \frac{dr}{d\tau}&=&\pm\sqrt{\mathcal{R}(r)} ,\label{r}\\
\Sigma \frac{d\theta}{d\tau}&=&\pm\sqrt{\Theta(\theta)} ,\label{th}\\
\Sigma \frac{d\phi}{d\tau}&=&\frac{a}{\Delta}\left({ E}(r^2+a^2)-a{ L_z}\right)-\left(a{ E}-\frac{{ L_z}}{\sin^2\theta}\right), \label{phiuch}
\end{eqnarray}
where $\tau$ is the affine parameter along the null geodesics and 
\begin{eqnarray}\label{06}
\mathcal{R}(r)&=&\left((r^2+a^2){E}-a{ L_z}\right)^2-\Delta ((a{ E}-{ L_z})^2+{ K}),\quad \\ 
\Theta(\theta)&=&{ K}-\left(\frac{{ L_z}^2}{\sin^2\theta}-a^2 { E}^2\right)\cos^2\theta.\label{theta0}
\end{eqnarray}
The constant ${K}$ is the separability constant related to the Carter constant $\mathcal{Q}$ through $\mathcal{Q}={K}+(aE-L_z)^2$. Carter constant appears as a conserved quantity associated with the hidden symmetry described by the second-rank Killing tensor. We introduce the impact parameters for the photons geodesics, which are constant along geodesics and defined in dimensionless form as follows  \cite{Chandrasekhar:1992}
\begin{eqnarray}
\xi=\frac{L_z}{E} \;\; \text{,} \;\; \eta=\frac{{K}}{E^2}.
\end{eqnarray}
Photons may get scattered, captured, or follow bound orbits around the black hole depending on the values of ($\xi, \eta$).  Because of the black hole rotation, photons can either co-rotate or counter-rotate along with the black hole, whose radii vary differently with black hole spin. At $\theta=\pi/2$, Carter's constant vanishes, and the photons follow the circular orbits with radii $r_p^{\pm}$, which can be determined by solving $Y=0$. Whereas for $\theta\neq \pi/2$ the Carter constant is positive definite and the photons follow the non-planar orbits with radii $r_p^-<r_p<r_p^+$. The photons following the spherical orbits of constant coordinate radii $r_p$ around the black hole are characterized by $\dot{r}=0$ and $\ddot{r}=0$. This results into the critical values of impact parameters ($\xi_{crit}, \eta_{crit}$) for the unstable orbits \cite{Kumar:2020owy}
\begin{align}
\xi_{crit}=&\frac{\left(a^2+r^2\right) \Delta '(r)-4 r \Delta (r)}{a \Delta '(r)},\nonumber\\
\eta_{crit}=&\frac{r^2 \left(8 \Delta (r) \left(2 a^2+r \Delta '(r)\right)-r^2 \Delta '(r)^2-16 \Delta (r)^2\right)}{a^2 \Delta '(r)^2}\label{CriImpPara},
\end{align}
where $'$ stands for the derivative with the radial coordinate $r$. Furthermore, these spherical photons orbits are the non-planar orbits that periodically cross the equatorial plane and construct a photon region around the black hole. As a result, the photons following the spherical orbits, beside having a motion along the $\phi$ direction also move along the $\theta$-direction.
For visualizing the black hole shadow, we consider a distant observer at position $(r_0, \theta_0)$. The coordinates ($X$,$Y$) define the observer image plane, such that the stereographic projection of the shadow from the celestial sphere to the image plane is defined as follow
\begin{align}
X=&\lim_{r_0\rightarrow\infty}\left(-r_0^2 \sin{\theta_0}\frac{d\phi}{d{r}}\right),\nonumber\\ Y=&\lim_{r_0\rightarrow\infty}\left(r_0^2\frac{d\theta}{dr}\right),\label{Celestial1}
\end{align} 
%%%%%%%%%%%%%%%%%%%%%%%%%%%%%%%%%%%%%%%%%%%%%%%%%%%%%%%%%%%%%%%%%%%%%%%%
For an asymptotically far observer, Eq.~(\ref{Celestial1}) leads to
\begin{align}
X=&-\xi_{crit}\csc\theta_0,\nonumber\\
Y=&\pm\sqrt{\eta_{crit}+a^2\cos^2\theta_0-\xi_{crit}^2\cot^2\theta_0}\ .\label{Celestial2}    
\end{align}
and satisfy
\begin{equation}
X^2+Y^2=\eta_{crit}+\xi_{crit}^2+a^2\cos^2\theta_0.\label{CelCoor}
\end{equation}
The parametric curve $Y$ vs $X$ delineates the shadow's boundary of the rotating EGB black hole. It is clear that the non-rotating black hole ($a=0$) cast a perfectly circular shadow silhouette. For $\theta_0\neq 0$ or $\pi $ the rotating black hole shadow shifts in the direction perpendicular to the black hole rotation and appears distorted, whereas for $\theta_0= 0,\pi $ shadows is centered at ($0,0$) and is perfectly circular for all values of $a$. The maximum off-center displacement of shadow appears for $\theta_0=\pi/2$. 

\begin{figure*}
	\begin{center}
	\begin{tabular}{c c}
		\includegraphics[scale=0.8]{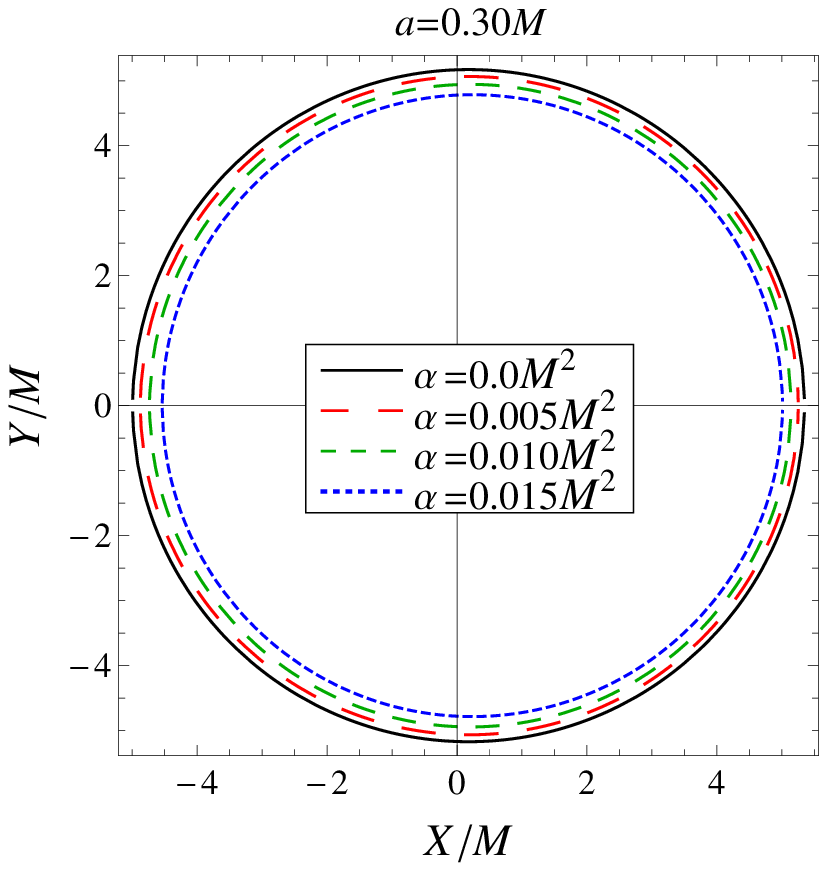}\hspace*{-0.5cm}&
		\includegraphics[scale=0.8]{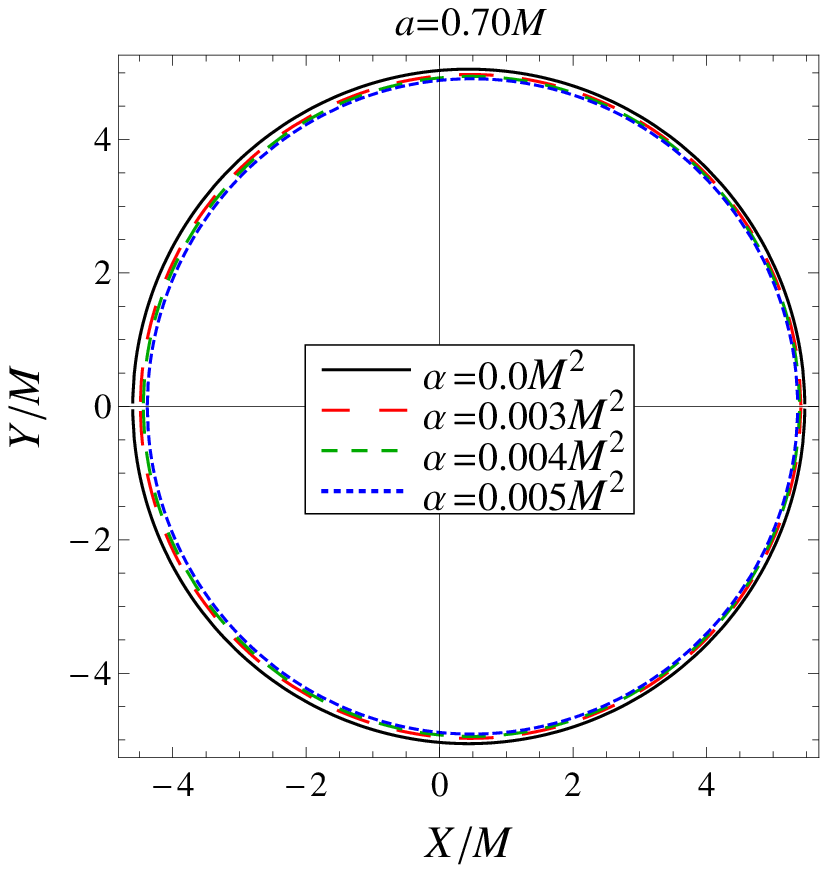}\\
		\includegraphics[scale=0.8]{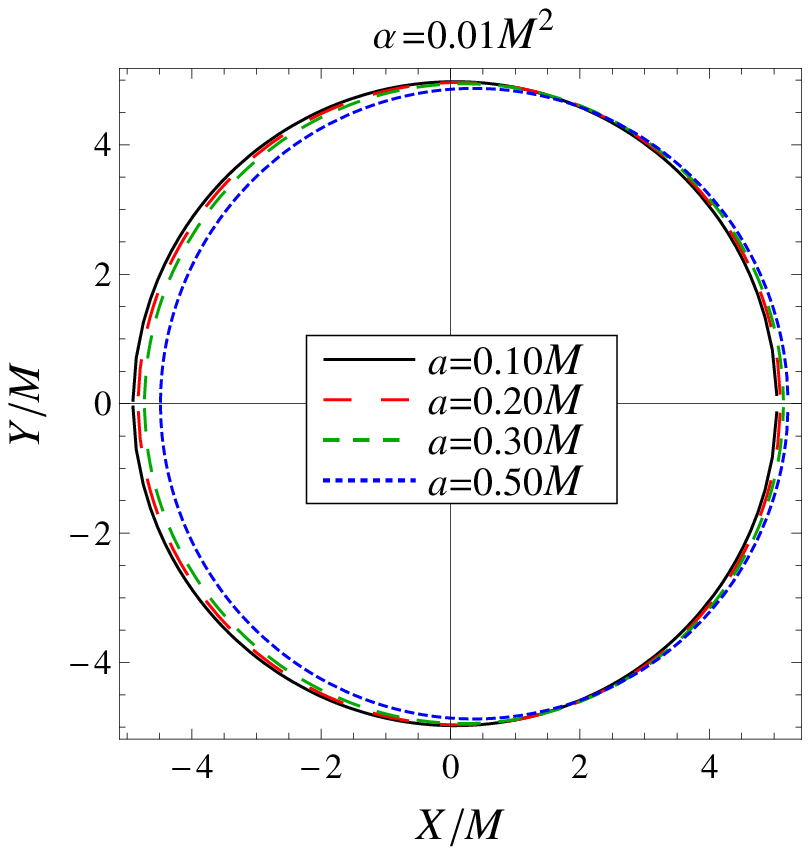}\hspace*{-0.5cm}&
		\includegraphics[scale=0.8]{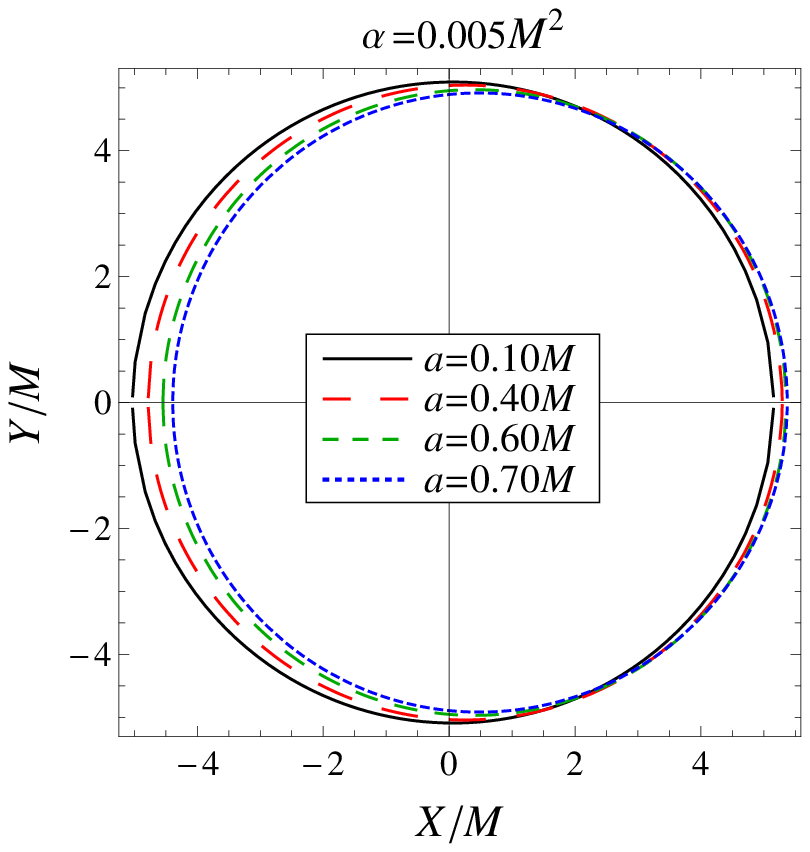}
	\end{tabular}
	\caption{Plot showing the rotating EGB black holes shadows with varying parameters $a$ and $\alpha$. Solid black curves in the upper panel are for the Kerr black holes. }
	\label{shadow}
	\end{center}
\end{figure*}
\begin{figure*}
	\begin{tabular}{c c c}
		\includegraphics[scale=0.6]{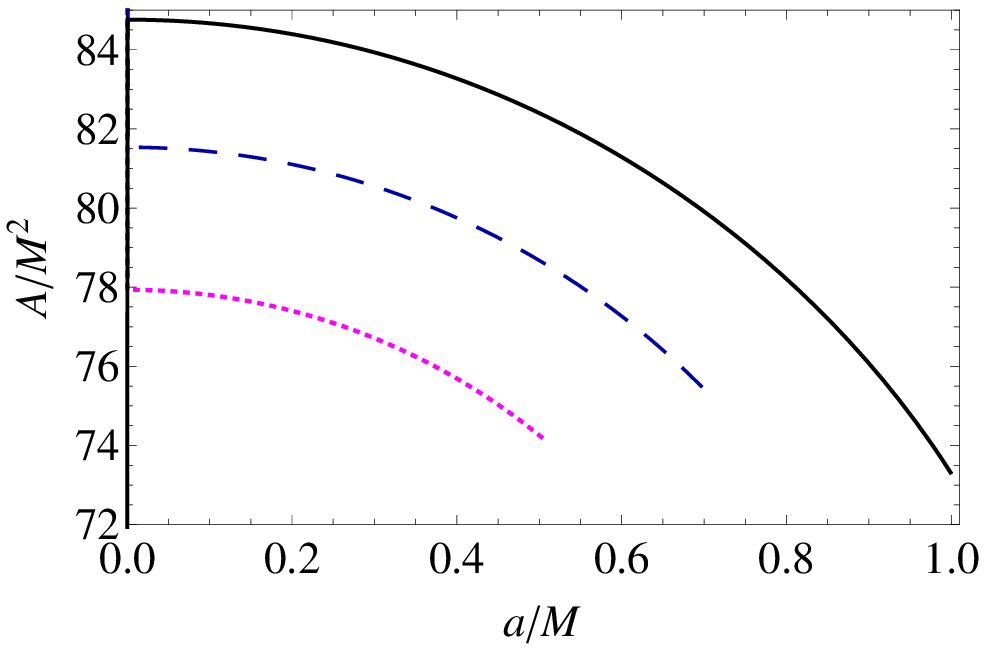}&
		\includegraphics[scale=0.6]{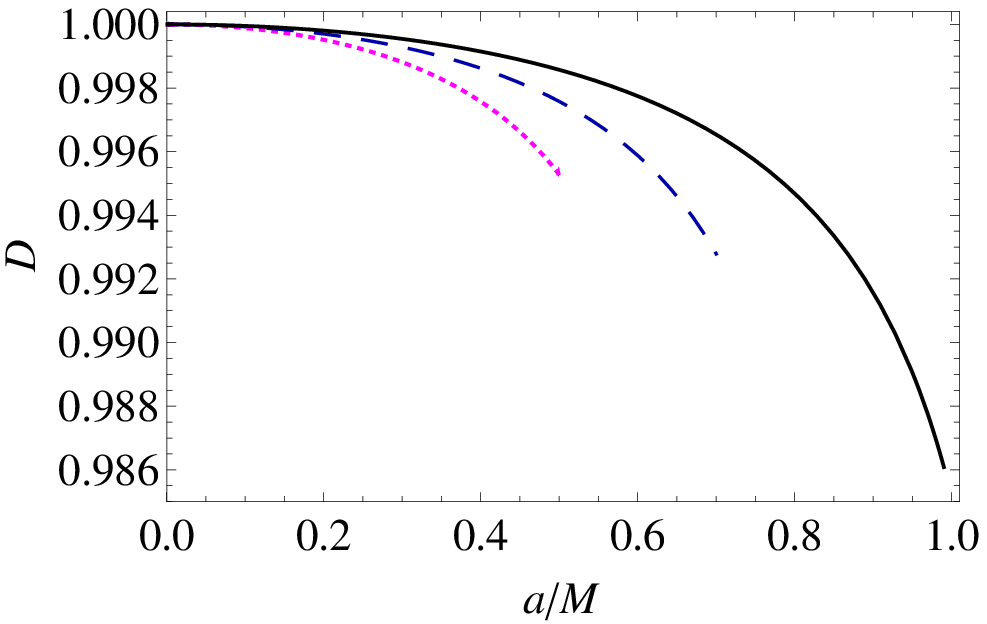}\\
		\includegraphics[scale=0.6]{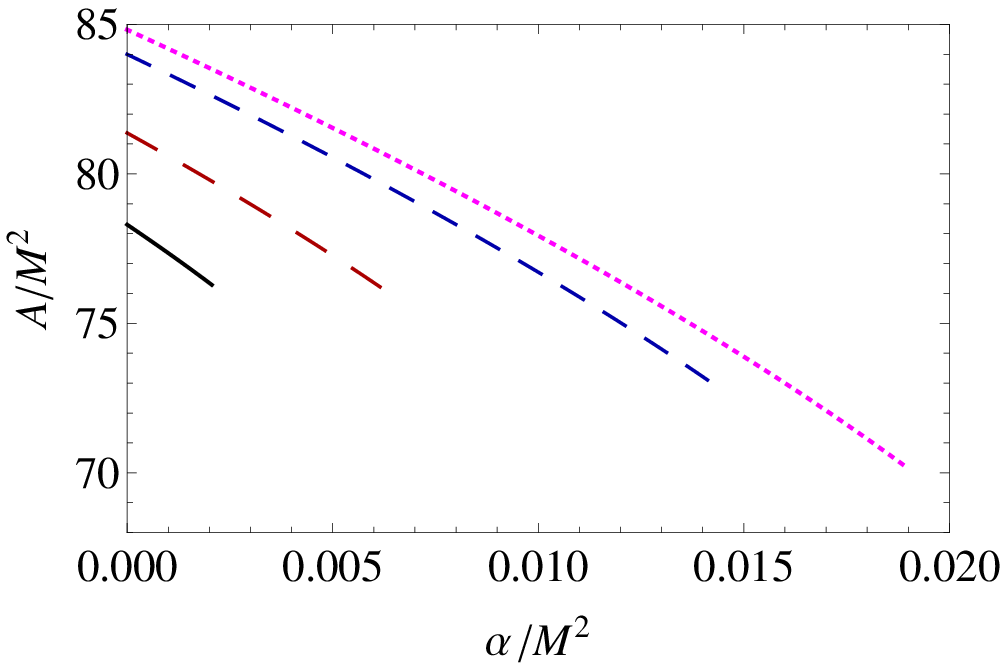}&
		\includegraphics[scale=0.6]{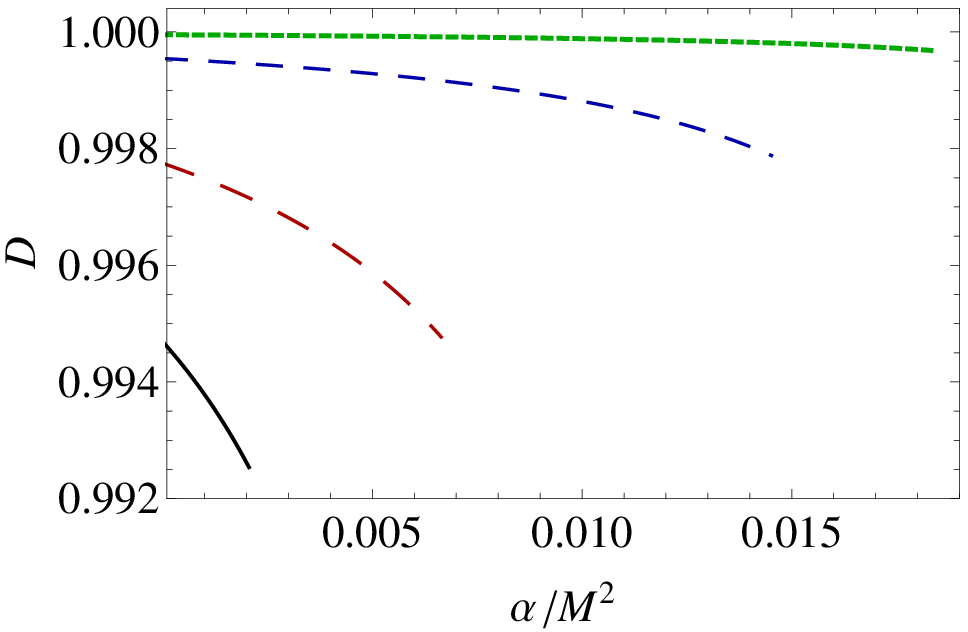}
	\end{tabular}
	\caption{\textit{Upper}: The observables $A$ and $D$ vs $a$ for $\alpha=0.0M^2$ (solid black curve), for $\alpha=0.005M^2$ (dashed blue curve), and for $\alpha=0.01M^2$ (dotted magenta curve). \textit{Bottom}: The observables $A$ and $D$ vs $\alpha$ for $a=0.0M$ (dotted magenta curve), for $a=0.1M$ (dotted green curve), for $a=0.3M$ (dashed blue curve), for $a=0.6M$ (long-dashed brown curve), and for $a=0.8M$ (solid black curve).}
	\label{obs}
\end{figure*}
\begin{figure*}
	\begin{center}
		\includegraphics[scale=0.9]{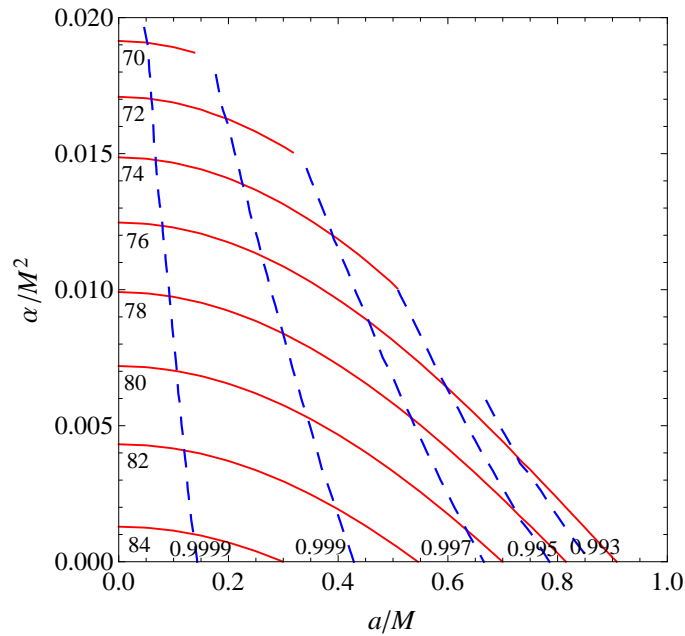}
	\end{center}
	\caption{Contour plots of the observables $A$ and $D$ in the plane $(a, \alpha)$ for the rotating EGB black holes. Each curve is labeled with the corresponding values of $A$ and  $D$. Solid red curves correspond to the $A$, and dashed blue curves are for the oblateness parameter $D$. }
	\label{parameterestimation}
\end{figure*}

In April 2019, the EHT collaboration using the VLBI technology unveiled the first-ever horizon-scaled image of the supermassive black hole M87* \cite{Akiyama:2019cqa, Akiyama:2019eap}. The M87* image shows powerful relativistic jets, which could be emerged from magnetohydrodynamic interactions between the accretion disk and the rotating black hole. Considering the orientation of these jets in M87*, the inclination angle (angle between the rotational axis and the line of sight) is estimated to be $17^o $ \cite{Walker:2018vam}. Hereafter, for our analysis of EGB black hole shadows, we will consider the inclination angle $\theta_0=17^o$. The rotating EGB black holes shadows with varying $a$ and $\alpha$ are depicted in Fig.~\ref{shadow}. It is evident that the shadow size decreases with increasing $\alpha$, such that the rotating EGB black hole shadows are smaller than the Kerr black hole shadows. Furthermore, the rotating black holes shadows are not perfectly circular. To characterize the shadow size and the deviation from the circularity, we introduce the shadow observables, namely, shadow area $A$ and oblateness $D$ as follows \cite{Kumar:2018ple, Tsupko:2017rdo}
\begin{equation}
A=2\int{Y dX}=2\int_{r_p^{-}}^{r_p^+}\left( Y \frac{dX}{dr_p}\right)dr_p,\label{Area}
\end{equation} 
\begin{equation}
D=\frac{X_r-X_l}{Y_t-Y_b},\label{Oblateness}
\end{equation}
where the points on the right, left, top, and bottom of the shadow boundary are designated by subscripts $r, l, t$ and $b$. The shadow boundary, defined by ($X, Y$), is a function of spherical photon orbits radii $r_p$. Interestingly, photons with different orbit radii construct the different parts of the shadow boundary. Therefore, as one moves up (down) along the $Y$-axis in the shadow image, one sees deeper (far away) from the black hole. The behavior of the shadow observables $A$ and $D$ with varying $a$ and $\alpha$ is shown in Fig.~\ref{obs}. The shadow area monotonically decreases with $\alpha$ and $a$, and the oblateness increases with increasing $a$. To estimate the rotating EGB black hole parameters, we make a contour plot of observables $A$ and $D$ as functions of $a$ and $\alpha$ in Fig.~\ref{parameterestimation}. Therein each solid red curve corresponds to constant values of $A$ and dashed blue curve to $D$. The intersection point of observables $A$ and $D$ determines the unique and precise values of the black hole parameters $a$ and $\alpha$. Hence, from Fig.~\ref{parameterestimation}, it is clear that for a given set of $4D$ EGB black hole shadow observables, $A$ and $D$, we can determine information about black hole spin and GB coupling parameter. 
%%%%%%%%%%%%%%%%%%%%%%%%%%%%%%%%%%%%%%%%%%%
\section{Constraints from the M87* shadow}
Although the M87* black hole shadow is found to be consistent with the Kerr black hole image as predicted in the general relativity, the non-Kerr black holes are also not ruled out. Very recently, the EHT collaboration team has set stringent constraints on the physical charges of a large variety of modified gravity black holes by using the M87* shadow  \cite{Kocherlakota:2021dcv}. The M87* shadow is of crescent shape with the circularity deviation $\Delta C\leq0.10$ (10\%) in terms of root-mean-square deviation from average shadow radius, axis ratio $\lesssim 4/3$, whereas the angular diameter $\theta_d$ is $42 \pm 3\mu as$ \cite{Akiyama:2019cqa,Akiyama:2019eap}. Here, we will model the M87* black hole as the rotating $4D$ EGB black hole and use the M87* shadow observables to place constraints on the black hole parameters. 

\begin{figure*}
	\begin{center}
		\includegraphics[scale=0.7]{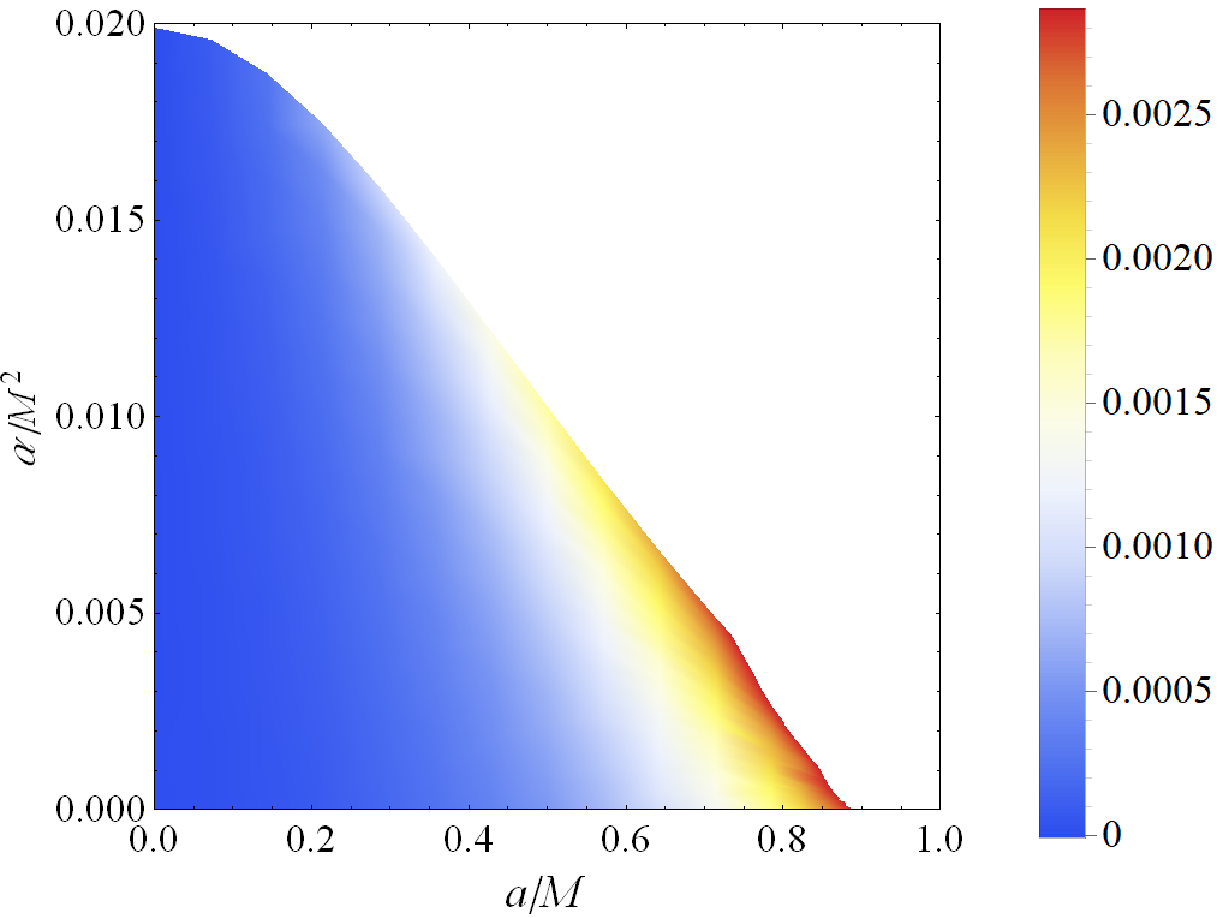}\\
		\includegraphics[scale=0.7]{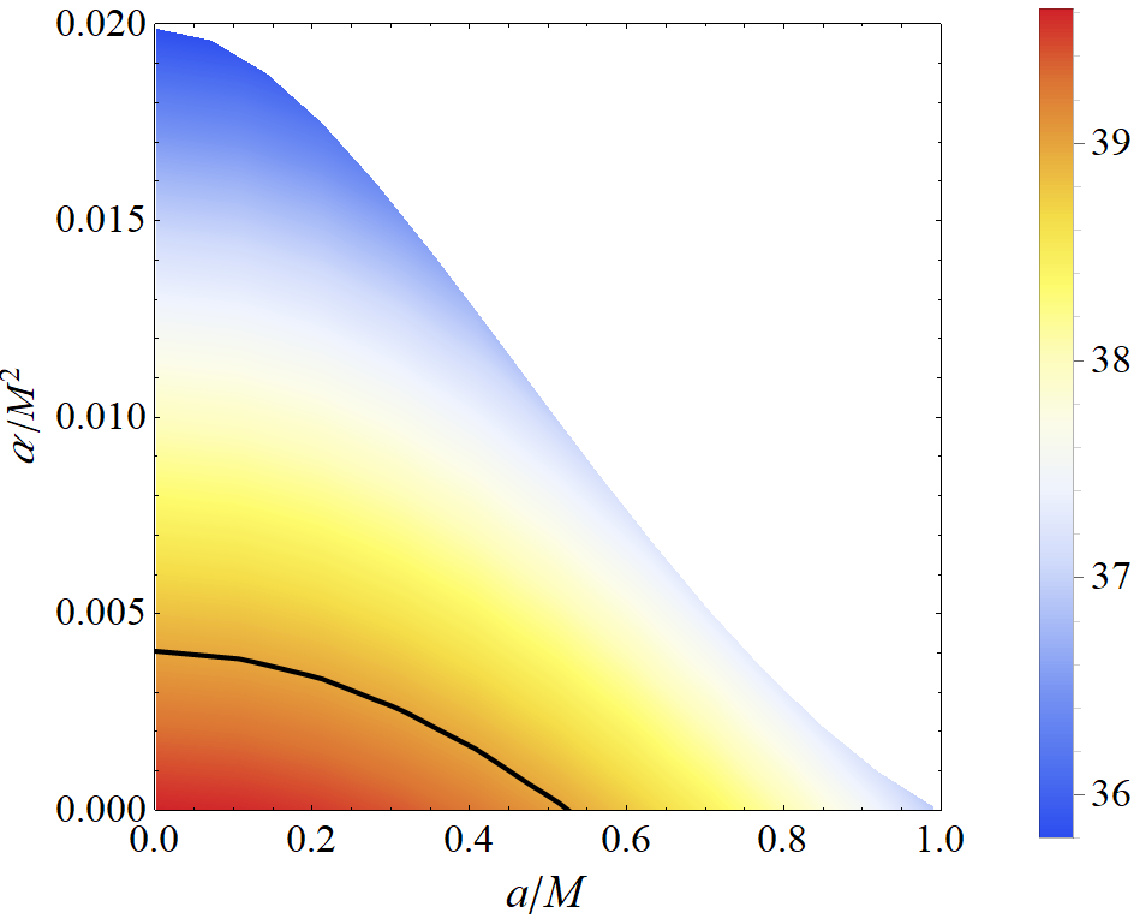}\\
		\includegraphics[scale=0.7]{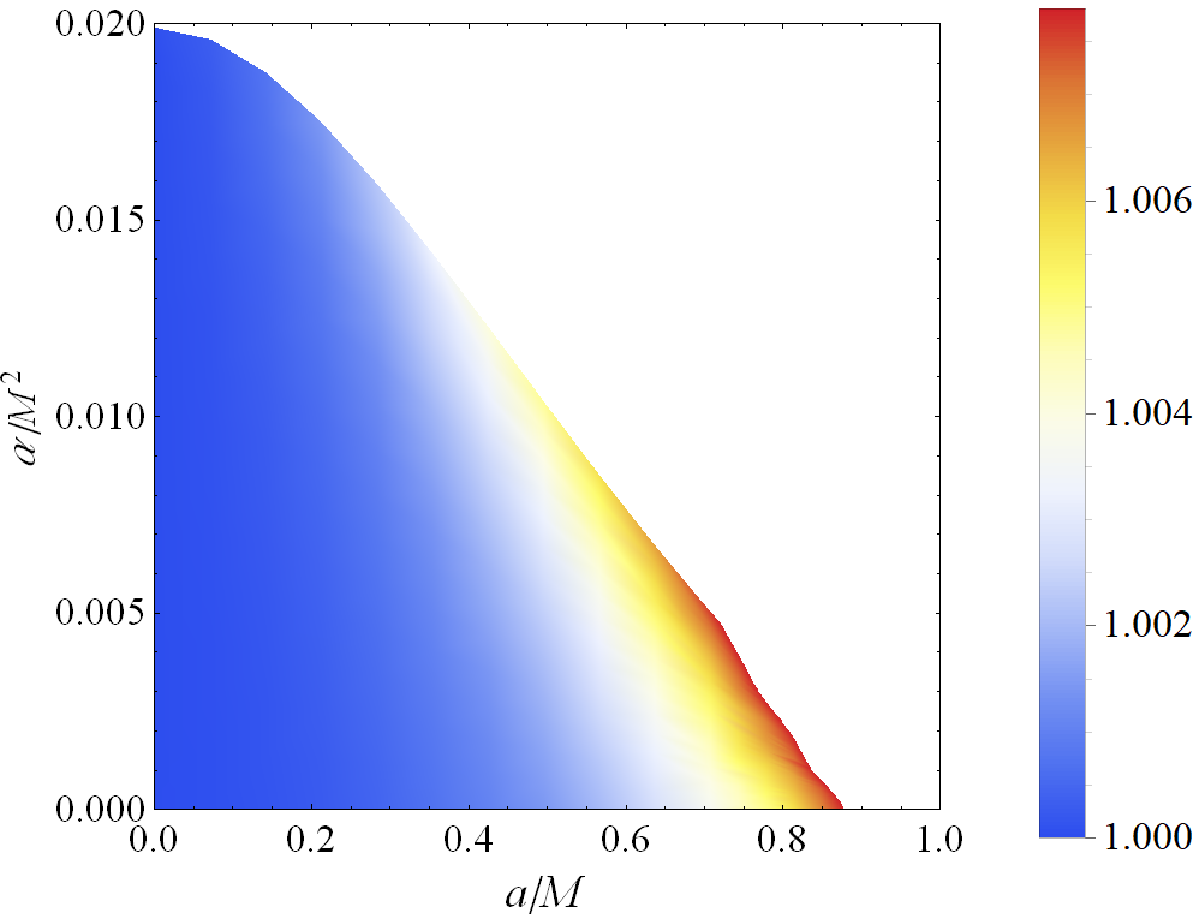}
	\caption{Circularity deviation observable $\Delta C$ (\textit{top}), the angular diameter $\theta_d$ (\textit{middle}), and axis ratio $D_x$ (\textit{bottom}) as a function of ($a,\alpha$) for the rotating EGB black holes. Black solid lines correspond to the M87* black hole shadow bounds $\theta_d=39\mu$as within the $1\sigma$ region, such that the region above the black line is excluded by the EHT bounds.}
	\label{M87}
	\end{center}
\end{figure*}

\begin{figure*}
	\begin{center}
		\includegraphics[scale=0.8]{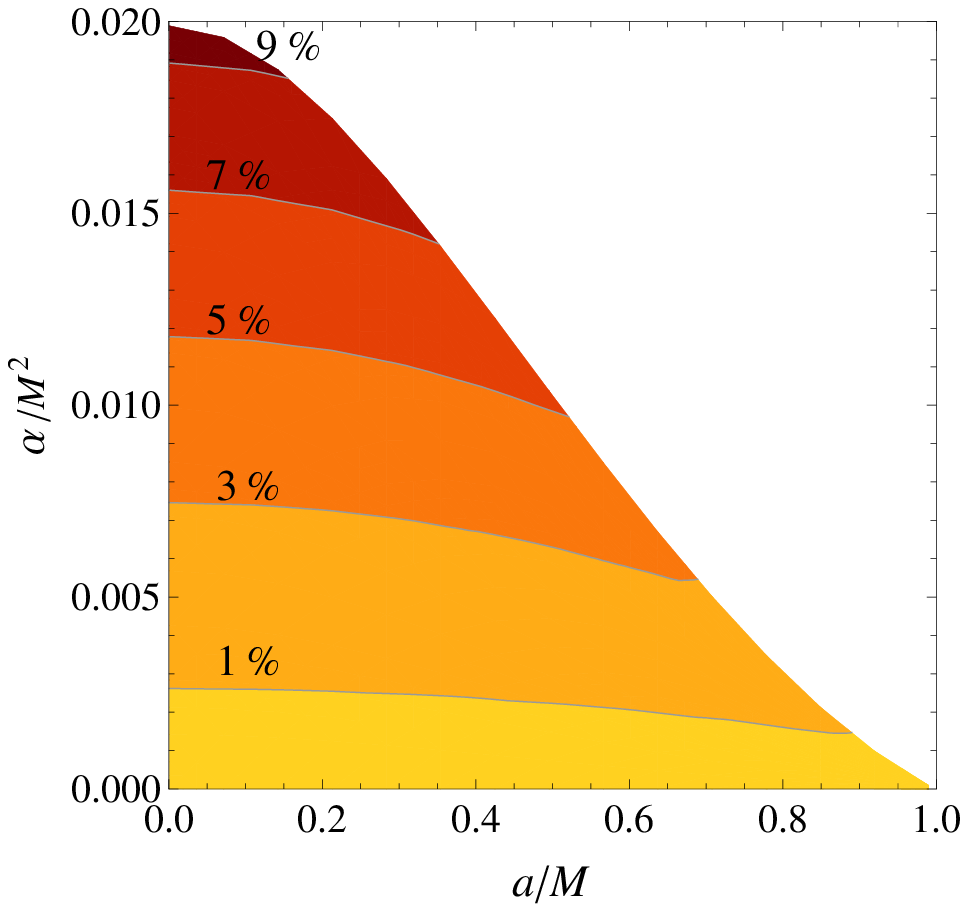}\\
		\caption{Relative difference between the shadow angular diameter size of a rotating EGB black hole and a Kerr black hole with the same mass. }
		\label{M87a}
	\end{center}
\end{figure*}

For this, we define the shadow boundary with polar coordinates ($R(\varphi),\varphi$) such that the origin is at the shadow center ($X_O,Y_O$). Figure (\ref{shadow}) infers that the rotating black hole shadow is always $\mathcal{Z}_2$ symmetric around $Y=0$. However, due to black hole rotation, the shadow center shifts from $X=0$, and as a result, the shadow is asymmetric along the $Y$ axis. It ascertains that the shadow center is $(X_O=|X_r + X_l|/2, Y_O=0)$, where $X_r$ and $X_l$ are the maximum and minimum abscissas of the shadow boundary in the image plane. The radial coordinate and polar angle of shadow boundary from its center reads as
\begin{equation}
R(\varphi)= \sqrt{(X-X_O)^2+(Y-Y_O)^2},\;\ \varphi\equiv \tan^{-1}\left(\frac{Y}{X-X_O}\right)\nonumber,
\end{equation}
whereas shadow average radius $\bar{R}$ is defined as \cite{Bambi:2019tjh}
\begin{eqnarray}
\bar{R}^2=\frac{1}{2\pi}\int_{0}^{2\pi} R^2(\varphi) d\varphi,
\end{eqnarray}
We define the dimensionless circularity deviation $\Delta C$ to quantifies the shadow deviation from a perfect circle as a measure of the root-mean-square deviation of $R(\varphi)$ from the shadow average radius  \cite{Johannsen:2010ru,Johannsen:2015qca,Bambi:2019tjh}
\begin{equation}
\Delta C=\frac{1}{\bar{R}}\sqrt{\frac{1}{2\pi}\int_0^{2\pi}\left(R(\varphi)-\bar{R}\right)^2d\varphi},
\end{equation}
clearly, for a circular shadow of spherically symmetric black hole $\Delta C=0$. Second observable is the shadow angular diameter $\theta_d$, which for a far distant observer, at a distance $r_0$ from the black hole, is defined as
\begin{equation}
\theta_d=2\frac{R_s}{r_0},\;\;\;\; R_s=\sqrt{A/\pi},
\end{equation}    
and the third observable is the axis ratio 
\begin{equation}
D_x=\frac{\Delta Y}{\Delta X},
\end{equation}
which is just the inverse of oblateness observable $D_x=1/D$.
We have calculated these three observable $\Delta C, \theta_d$, and $D_x$ for the rotating $4D$ EGB black hole with $M=6.5\times 10^9 M_{\odot}$ and $r_0=16.8$ Mpc and plotted them in Fig.~\ref{M87}. The EHT bound for the M87* black hole shadow angular diameter $\theta_d=39\,\mu$as within the $1\sigma$ region, shown as the black solid line, constrained the $a$ and $\alpha$. The shadow angular size for the non-rotating extremal EGB black hole with $a=0,\alpha=0.019894367M^2$ is $\theta_d=35.7888\mu$as and for extremal Kerr black hole with $a=M, \alpha=0$ is $\theta_d=36.8632\mu$as. For comparison, at the inclination angle $\theta_o=90^o$, the extremal Kerr black hole shadow angular size is $37.3534\mu$as. The relative difference in shadow angular diameter $\delta\theta_d\equiv (\theta_d|_{Kerr}-\theta_d|_{EGB})/\theta_d|_{Kerr}$ is shown in Fig.~\ref{M87a}. Clearly, $\delta\theta_d\leq 17\%$ and thus it is consistent with the Psaltis \textit{et al.} \cite{Psaltis:2020lvx} findings. Furthermore, the axis ratio and the circularity deviation for the M87* black hole shadow allow all parameter space of the EGB black hole.

\section{Conclusions}
The underlying theory of gravity in the extreme-field regime is currently unknown, and insights into it are likely to be gained through observations. In this paper, we have investigated the rotating $4D$ EGB black hole. The EGB gravity theory has been of great interest and importance due to second-order field equations and being free from ghost instabilities. The rotating black holes possess two distinct horizons that eventually merge to form the degenerate horizon for the extremal values of the GB coupling parameter $\alpha=\alpha_E$. The null geodesics equations of motion were obtained in the first-order differential form, and the analytical expressions of the critical impact parameters for unstable spherical photon orbits are derived. The shadow contours are drawn for the rotating black hole for $\theta_0=17^o$ inclination angle and compared with those for the Kerr black hole. The rotating black hole shadows deviate from the circularity. The shadow observables $A$ and $D$ are calculated, and it is shown that the EGB black hole shadows are smaller and more distorted than those for the Kerr black hole. Furthermore, these observables are used to estimate the black hole parameters. We modeled the M87* black hole as the rotating $4D$ EGB black hole and used the deduced shadow observables $\Delta C$, $\theta_d$, and $D_x$ for the M87* to constrain the EGB black hole parameters. We have found that only $\theta_d$ within $1\sigma$ region placed stringent bound the EGB parameters. Whereas $\Delta C$ and $D_x$ allows all parameter space. The constraints deduced for $\theta_0=17^o$ are weaker than those deduced for $\theta_0=90^o$ in Ref.~\cite{Kumar:2020owy}. However, it is important to account for systematic uncertainty when identifying observable shadow characteristics like the emission ring and center brightness depression, especially when using low-resolution data, to gravitational qualities like the size and shape of the critical curve. Future observations utilizing an improved ground or space-based array might significantly reduce these systematic errors, and we anticipate better constraints on the GB coupling parameter.

\section{Acknowledgments}

S.G.G. would like to thank DST INDO-SA bilateral project DST/INT/South Africa/P-06/2016, SERB-DST for the ASEAN project IMRC/AISTDF/CRD/2018/000042 and also IUCAA, Pune for the hospitality while this work was being done. R.K. would like to thanks UKZN and NRF for the post-doctoral fellowship.

%Non BiBTeX users can list down their references as:


\begin{thebibliography}{10}
%\cite{Lovelock:1972vz}
\bibitem{Lovelock:1972vz}
D.~Lovelock,
J.\ Math.\ Phys.\  {\bf 13} 874 (1972).
%doi:10.1063/1.1666069

%\cite{Lanczos:1938sf}
\bibitem{Lanczos:1938sf}
C.~Lanczos,
Annals Math.\  {\bf 39} 842 (1938).
%doi:10.2307/1968467


%\cite{Lovelock:1971yv}
\bibitem{Lovelock:1971yv}
D.~Lovelock,
J.\ Math.\ Phys.\  {\bf 12}  498 (1971).
%doi:10.1063/1.1665613

%\cite{Boulware:1985wk}
\bibitem{Boulware:1985wk}  D.G.~Boulware and S.~Deser, Phys. Rev. Lett. {\bf 55}, 2656 (1985); 
J.T.~Wheeler, Nucl. Phys. B {\bf 268}, 737 (1986). 

\bibitem {egb}
%\cite{Nojiri:2001aj}
S.~Nojiri and S.~D.~Odintsov,
%``Anti-de Sitter black hole thermodynamics in higher derivative gravity and new confining deconfining phases in dual CFT,''
Phys.\ Lett.\ B {\bf 521} 87 (2001);
Erratum: [Phys.\ Lett.\ B {\bf 542} 301 (2002)];
Y. M. Cho and I. P. Neupane, Phys. Rev. D {\bf 66}, 024044 (2002); 
M. Cvetic, S. Nojiri and S. D. Odintsov, Nucl. Phys. {\bf B 628}, 295 (2002); 
R. G. Cai, Phys. Rev. D {\bf 65}, 084014 (2002); 
I. P. Neupane, Phys. Rev. D {\bf 67}, 061501(R) (2003); {\bf 69}, 084011 (2004);  
A. Padilla, Class. Quant. Grav. {\bf 20}, 3129 (2003); 
N. Deruelle, J. Katz, and S. Ogushi, Class. Quant. Grav. {\bf 21}, 1971 (2004); 
M. H. Dehghani, Phys. Rev. D {\bf 69}, 064024 (2004);  
R. G. Cai and Q. Guo, Phys. Rev. D {\bf 69}, 104025 (2004); 
T. Torii and H. Maeda, Phys. Rev. D {\bf 71}, 124002 (2005); 
M. H. Dehghani and R. B. Mann, Phys. Rev. D {\bf 72}, 124006 (2005); 
M. H. Dehghani and S. H. Hendi, Phys. Rev. D {\bf 73}, 084021 (2006); 
M. H. Dehghani, G. H. Bordbar, and M. Shamirzaie, Phys. Rev. D {\bf 74}, 064023 (2006).



\bibitem{ghosh} S.~Jhingan and S.~G.~Ghosh,
%``Inhomogeneous dust collapse in D-5 Einstein-Gauss-Bonnet gravity,''
Phys.\ Rev.\ D {\bf 81}, 024010 (2010);
S.~G.~Ghosh, M.~Amir and S.~D.~Maharaj,
%``Quintessence background for 5D Einstein?Gauss?Bonnet black holes,''
Eur.\ Phys.\ J.\ C {\bf 77}, 530 (2017);
S.~G.~Ghosh,
%``Noncommutative geometry inspired Einstein?Gauss?Bonnet black holes,''
Class.\ Quant.\ Grav.\  {\bf 35}, 085008 (2018). 
\bibitem{egb2} 
S.~Mignemi and N.~R.~Stewart,
%``Charged black holes in effective string theory,''
Phys.\ Rev.\ D {\bf 47}, 5259 (1993);
%\cite{Kanti:1995vq}
P.~Kanti, N.~E.~Mavromatos, J.~Rizos, K.~Tamvakis and E.~Winstanley,
%``Dilatonic black holes in higher curvature string gravity,''
Phys.\ Rev.\ D {\bf 54}, 5049 (1996);
%\cite{Alexeev:1996vs}
S.~O.~Alexeev and M.~V.~Pomazanov,
%``Black hole solutions with dilatonic hair in higher curvature gravity,''
Phys.\ Rev.\ D {\bf 55}, 2110 (1997);
%\cite{Torii:1996yi}
T.~Torii, H.~Yajima and K.~i.~Maeda,
%``Dilatonic black holes with Gauss-Bonnet term,''
Phys.\ Rev.\ D {\bf 55}, 739 (1997);
%\cite{Konoplya:2004xx}
R.~Konoplya,
%``Quasinormal modes of the charged black hole in Gauss-Bonnet gravity,''
Phys.\ Rev.\ D {\bf 71}, 024038 (2005);
%\cite{Kleihaus:2011tg}
B.~Kleihaus, J.~Kunz and E.~Radu,
%``Rotating Black Holes in Dilatonic Einstein-Gauss-Bonnet Theory,''
Phys.\ Rev.\ Lett.\  {\bf 106}, 151104 (2011);
%\cite{Maselli:2015tta}
A.~Maselli, P.~Pani, L.~Gualtieri and V.~Ferrari,
%``Rotating black holes in Einstein-Dilaton-Gauss-Bonnet gravity with finite coupling,''
Phys.\ Rev.\ D {\bf 92}, 083014 (2015).


%\cite{Sotiriou:2013qea}
\bibitem{Sotiriou:2013qea}
T.~P.~Sotiriou and S.~Y.~Zhou,
%``Black hole hair in generalized scalar-tensor gravity,''
Phys. Rev. Lett. \textbf{112}, 251102 (2014).

%\cite{Sotiriou:2014pfa}
\bibitem{Sotiriou:2014pfa}
T.~P.~Sotiriou and S.~Y.~Zhou,
%``Black hole hair in generalized scalar-tensor gravity: An explicit example,''
Phys. Rev. D \textbf{90}, 124063 (2014).

%\cite{Doneva:2017bvd}
\bibitem{Doneva:2017bvd}
D.~D.~Doneva and S.~S.~Yazadjiev,
%``New Gauss-Bonnet Black Holes with Curvature-Induced Scalarization in Extended Scalar-Tensor Theories,''
Phys. Rev. Lett. \textbf{120}, 131103 (2018).

%\cite{Cunha:2019dwb}
\bibitem{Cunha:2019dwb}
P.~V.~P.~Cunha, C.~A.~R.~Herdeiro and E.~Radu,
%``Spontaneously Scalarized Kerr Black Holes in Extended Scalar-Tensor\textendash{}Gauss-Bonnet Gravity,''
Phys. Rev. Lett. \textbf{123}, 011101 (2019).

\bibitem{Tomozawa:2011gp}
%\cite{Tomozawa:2011gp}
Y.~Tomozawa,
%``Quantum corrections to gravity,''
arXiv:1107.1424 [gr-qc].


%\cite{Cognola:2013fva}
\bibitem{Cognola:2013fva} 
G.~Cognola, R.~Myrzakulov, L.~Sebastiani and S.~Zerbini,
%``Einstein gravity with Gauss-Bonnet entropic corrections,''
Phys.\ Rev.\ D {\bf 88}, 024006 (2013).


%\cite{Glavan:2019inb}
\bibitem{Glavan:2019inb}
D.~Glavan and C.~Lin,
%``Einstein-Gauss-Bonnet gravity in 4-dimensional space-time,''
Phys.\ Rev.\ Lett.\  {\bf 124}, 081301 (2020).


%\cite{Konoplya:2020qqh}
\bibitem{Konoplya:2020qqh}
R.~A.~Konoplya and A.~Zhidenko,
%``Black holes in the four-dimensional Einstein-Lovelock gravity,''
Phys. Rev. D \textbf{101}, 084038 (2020).


%\cite{Casalino:2020kbt}
\bibitem{Casalino:2020kbt} 
A.~Casalino, A.~Colleaux, M.~Rinaldi and S.~Vicentini,
%``Regularized Lovelock gravity,''
Phys. Dark Univ. \textbf{31}, 100770 (2021).

\bibitem{Hennigar:2020lsl}
R.~A.~Hennigar, D.~Kubiznak, R.~B.~Mann and C.~Pollack,
%``On Taking the $D\to 4$ limit of Gauss-Bonnet Gravity: Theory and Solutions,''
JHEP \textbf{07}, 027 (2020).


%\cite{Ma:2020ufk}
\bibitem{Ma:2020ufk}
L.~Ma and H.~Lu,
%``Vacua and Exact Solutions in Lower-$D$ Limits of EGB,''
Eur. Phys. J. C \textbf{80}, 1209 (2020).

\bibitem{Ai:2020peo}
W.~Ai,
%``A note on the novel 4D Einstein-Gauss-Bonnet gravity,''
Commun. Theor. Phys. \textbf{72}, 095402 (2020).


\bibitem{Shu:2020cjw}
F.~Shu,
%``Vacua in novel 4D Eisntein-Gauss-Bonnet Gravity: pathology and instability,''
Phys. Lett. B \textbf{811}, 135907 (2020).


\bibitem{Gurses:2020ofy}
M.~Gurses, T.~C.~Sisman and B.~Tekin,
%``Is there a novel Einstein-Gauss-Bonnet theory in four dimensions?,''
Eur. Phys. J. C \textbf{80} 647 (2020).

\bibitem{Mahapatra:2020rds}
S.~Mahapatra,
%``A note on the total action of $4D$ Gauss-Bonnet theory,''
Eur. Phys. J. C \textbf{80}, 992 (2020).


\bibitem{Lu:2020iav}
H.~Lu and Y.~Pang,
%``Horndeski Gravity as $D\rightarrow4$ Limit of Gauss-Bonnet,''
Phys. Lett. B \textbf{809}, 135717 (2020).

\bibitem{Kobayashi:2020wqy}
T.~Kobayashi,
%``Effective scalar-tensor description of regularized Lovelock gravity in four dimensions,''
JCAP \textbf{07}, 013 (2020).

%\cite{Arrechea:2020evj}
\bibitem{Arrechea:2020evj}
J.~Arrechea, A.~Delhom and A.~Jiménez-Cano,
%``Yet another comment on four-dimensional Einstein-Gauss-Bonnet gravity,''
Chin. Phys. C \textbf{45}, 013107 (2021).

%\cite{Aoki:2020lig}
\bibitem{Aoki:2020lig}
K.~Aoki, M.~A.~Gorji and S.~Mukohyama,
%``A consistent theory of $D\rightarrow 4$ Einstein-Gauss-Bonnet gravity,''
Phys. Lett. B \textbf{810}, 135843 (2020).


%\cite{Mann:1992ar}
\bibitem{Mann:1992ar}
R.~B.~Mann and S.~F.~Ross,
%``The D ---\ensuremath{>} 2 limit of general relativity,''
Class. Quant. Grav. \textbf{10}, 1405 (1993).

%\cite{Fernandes:2020nbq}
\bibitem{Fernandes:2020nbq}
P.~G.~S.~Fernandes, P.~Carrilho, T.~Clifton and D.~J.~Mulryne,
%``Derivation of Regularized Field Equations for the Einstein-Gauss-Bonnet Theory in Four Dimensions,''
Phys. Rev. D \textbf{102}, 024025 (2020).


%\cite{Fernandes:2020rpa}
\bibitem{Fernandes:2020rpa} 
P.~G.~S.~Fernandes,
%``Charged Black Holes in AdS Spaces in $4D$ Einstein Gauss-Bonnet Gravity,''
Phys. Lett. B \textbf{805}, 135468 (2020).


\bibitem{Singh:2020nwo}
D.~V.~Singh, S.~G.~Ghosh and S.~D.~Maharaj,
%``Clouds of string in the novel $4D$ Einstein-Gauss-Bonnet black holes,''
Phys. Dark Univ. \textbf{30}, 100730 (2020).

%\cite{Wei:2020ght}
\bibitem{Wei:2020ght} 
S.~W.~Wei and Y.~X.~Liu,
%``Testing the nature of Gauss-Bonnet gravity by four-dimensional rotating black hole shadow,''
Eur. Phys. J. Plus \textbf{136}, 436 (2021).

%\cite{Kumar:2020owy}
\bibitem{Kumar:2020owy} 
R.~Kumar and S.~G.~Ghosh,
%``Rotating black holes in the   $4D$ Einstein-Gauss-Bonnet gravity,''
JCAP \textbf{07}, 053 (2020).

%\cite{Ghosh:2020vpc}
\bibitem{Ghosh:2020vpc} 
S.~G.~Ghosh and S.~D.~Maharaj,
%``Radiating black holes in the   4D Einstein-Gauss-Bonnet gravity,''
Phys. Dark Univ. \textbf{30}, 100687 (2020).

%\cite{Ghosh:2020syx}
\bibitem{Ghosh:2020syx}
S.~G.~Ghosh and R.~Kumar,
%``Generating black holes in $4D$ Einstein-Gauss-Bonnet gravity,''
Class. Quant. Grav. \textbf{37},  245008 (2020).

\bibitem{Kumar:2020xvu}
A.~Kumar and S.~G.~Ghosh,
%``Hayward black holes in the novel $4D$ Einstein-Gauss-Bonnet gravity,''
arXiv:2004.01131 [gr-qc].

%\cite{Kumar:2020uyz}
\bibitem{Kumar:2020uyz}
A.~Kumar and R.~Kumar,
%``Bardeen black holes in the novel $4D$ Einstein-Gauss-Bonnet gravity,''
arXiv:2003.13104 [gr-qc].


\bibitem{Islam:2020xmy}
S.~U.~Islam, R.~Kumar and S.~G.~Ghosh,
%``Gravitational lensing by black holes in the $4D$ Einstein-Gauss-Bonnet gravity,''
JCAP \textbf{09}, 030 (2020).

%\cite{Heydari-Fard:2020sib}
\bibitem{Heydari-Fard:2020sib}
M.~Heydari-Fard, M.~Heydari-Fard and H.~R.~Sepangi,
%``Bending of light in novel 4$D$ Gauss-Bonnet-de Sitter black holes by Rindler-Ishak method,''
EPL \textbf{133}, 50006 (2021).

%\cite{Jin:2020emq}
\bibitem{Jin:2020emq}
X.~H.~Jin, Y.~X.~Gao and D.~J.~Liu,
%``Strong gravitational lensing of a 4-dimensional Einstein-Gauss-Bonnet black hole in homogeneous plasma,''
Int. J. Mod. Phys. D \textbf{29}, 2050065 (2020).

%\cite{Kumar:2020sag}
\bibitem{Kumar:2020sag}
R.~Kumar, S.~U.~Islam and S.~G.~Ghosh,
%``Gravitational lensing by Charged black hole in regularized $4D$ Einstein-Gauss-Bonnet gravity,''
Eur. Phys. J. C \textbf{80}, 1128 (2020).


%\cite{Akiyama:2019cqa}
\bibitem{Akiyama:2019cqa} 
K.~Akiyama {\it et al.},
%``First M87 Event Horizon Telescope Results. I. The Shadow of the Supermassive Black Hole,''
Astrophys.\ J.\  {\bf 875}, L1 (2019).


%\cite{Akiyama:2019eap}
\bibitem{Akiyama:2019eap} 
K.~Akiyama {\it et al.},
%``First M87 Event Horizon Telescope Results. VI. The Shadow and Mass of the Central Black Hole,''
Astrophys.\ J.\  {\bf 875}, L6 (2019).


\bibitem{Psaltis:2020lvx}
D.~Psaltis \textit{et al.},
%``Gravitational Test Beyond the First Post-Newtonian Order with the Shadow of the M87 BH,''
Phys. Rev. Lett. \textbf{125}, 141104 (2020).

%\cite{Psaltis:2008bb}
%\bibitem{Psaltis:2008bb}
%D.~Psaltis,
%``Probes and Tests of Strong-Field Gravity with Observations in the Electromagnetic Spectrum,''
%Living Rev.\ Rel.\  {\bf 11}, 9 (2008).

%\cite{Mizuno:2018lxz}
\bibitem{Mizuno:2018lxz}
Y.~Mizuno, \textit{et al.},
%``The Current Ability to Test Theories of Gravity with Black Hole Shadows,''
Nature Astron. \textbf{2}, 585 (2018).

%\cite{Vincent:2020dij}
\bibitem{Vincent:2020dij}
F.~H.~Vincent, M.~Wielgus, M.~A.~Abramowicz, E.~Gourgoulhon, J.~P.~Lasota, T.~Paumard and G.~Perrin,
%``Geometric modeling of M87* as a Kerr black hole or a non-Kerr compact object,''
Astron. Astrophys. \textbf{646}, A37 (2021).

%\cite{Junior:2021atr}
\bibitem{Junior:2021atr}
H.~C.~D.~Lima, Junior., L.~B.~Crispino, P.~P.~Cunha and C.~R.~Herdeiro,
%``Can different black holes cast the same shadow?,''
Phys. Rev. D \textbf{103}, 084040 (2021).

\bibitem{Newman:1965tw}
E.~Newman and A.~Janis,
%``Note on the Kerr spinning particle metric,''
J.\ Math.\ Phys.\  \textbf{6}, 915 (1965).

%\cite{Contreras:2021yxe}
\bibitem{Contreras:2021yxe}
E.~Contreras, J.~Ovalle and R.~Casadio,
%``Gravitational decoupling for axially symmetric systems and rotating black holes,''
Phys. Rev. D \textbf{103}, 044020 (2021).

\bibitem{Azreg-Ainou:2014pra} 
M.~Azreg-A{\"i}nou,
%``Generating rotating regular black hole solutions without complexification,''
Phys.\ Rev.\ D {\bf 90}, 064041 (2014).

\bibitem{Azreg-Ainou:2014aqa} 
M.~Azreg-A{\"i}nou,
%``From static to rotating to conformal static solutions: Rotating imperfect fluid wormholes with(out) electric or magnetic field,''
Eur.\ Phys.\ J.\ C {\bf 74}, 2865 (2014).


\bibitem{Johannsen:2011dh}
T.~Johannsen and D.~Psaltis,
%``A Metric for Rapidly Spinning Black Holes SuiTable for Strong-Field Tests of the No-Hair Theorem,''
Phys. Rev. D \textbf{83}, 124015 (2011).

%\cite{Bambi:2013ufa}
\bibitem{Bambi:2013ufa}
C.~Bambi and L.~Modesto,
%``Rotating   black holes,''
Phys.\ Lett.\ B \textbf{721}, 329 (2013).

\bibitem{Ghosh:2014pba}
S.~G.~Ghosh,
%``A nonsingular rotating black hole,''
Eur. Phys. J. C \textbf{75}, 532 (2015).

\bibitem{Moffat:2014aja} 
J.~W.~Moffat,
%``Black Holes in Modified Gravity (MOG),''
Eur.\ Phys.\ J.\ C {\bf 75}, 175 (2015).

\bibitem{Kumar:2020hgm}
R.~Kumar, S.~G.~Ghosh and A.~Wang,
%``Gravitational deflection of light and shadow cast by rotating Kalb-Ramond black holes,''
Phys. Rev. D \textbf{101}, 104001 (2020).

\bibitem{Kumar:2017qws}
R.~Kumar and S.~G.~Ghosh,
%``Rotating black hole in Rastall theory,''
Eur. Phys. J. C \textbf{78}, 750 (2018).

%\cite{Kerr:1963ud}
\bibitem{Kerr:1963ud} 
R.~P.~Kerr,
%``Gravitational field of a spinning mass as an example of algebraically special metrics,''
Phys.\ Rev.\ Lett.\  {\bf 11}, 237 (1963).

%\cite{Cai:2009ua}
\bibitem{Cai:2009ua} 
R.~G.~Cai, L.~M.~Cao and N.~Ohta,
%``Black Holes in Gravity with Conformal Anomaly and Logarithmic Term in Black Hole Entropy,''
JHEP {\bf 1004}, 082 (2010);
%\cite{Cai:2014jea}
R.~G.~Cai,
%``Thermodynamics of Conformal Anomaly Corrected Black Holes in AdS Space,''
Phys.\ Lett.\ B {\bf 733}, 183 (2014).

%\cite{Kehagias:2009is}
\bibitem{Kehagias:2009is}
A.~Kehagias and K.~Sfetsos,
%``The Black hole and FRW geometries of non-relativistic gravity,''
Phys. Lett. B \textbf{678}, 123 (2009).

\bibitem{Synge:1966} J.~L.~Synge,
%``The escape of photons from gravitationally intense stars,''
Mon.\ Not.\ R.\ Astron.\ Soc.\  {\bf 131}, 463 (1966).

\bibitem{Luminet:1979} J. P. Luminet, Astron. Astrophys. {\bf 75}, 228 (1979).

\bibitem{Bardeen} J. M. Bardeen, \textit{ Black Holes}, Edited
by C. DeWitt and B. S. DeWitt (Gordon and Breach,
New York, 1973, p. 215).

%\cite{Chandrasekhar:1992} 
\bibitem {Chandrasekhar:1992}
S.~ Chandrasekhar, 
{\it The Mathematical Theory of Black Holes} (Oxford University Press, New York, 1992).

%\cite{CraigWalker:2018vam}
\bibitem{Walker:2018vam}
R.~Craig Walker, P.~E.~Hardee, F.~B.~Davies, C.~Ly and W.~Junor,
%``The Structure and Dynamics of the Subparsec Jet in M87 Based on 50 VLBA Observations over 17 Years at 43 GHz,''
Astrophys. J. \textbf{855}, 128 (2018).

%\cite{Kumar:2018ple}
\bibitem{Kumar:2018ple} 
R.~Kumar and S.~G.~Ghosh,
%``Black hole parameters estimation from its shadow,''
Astrophys.\ J.\  {\bf 892}, 78 (2020).

%\cite{Tsupko:2017rdo}
\bibitem{Tsupko:2017rdo} 
O.~Y.~Tsupko,
%``Analytical calculation of black hole spin using deformation of the shadow,''
Phys.\ Rev.\ D {\bf 95}, 104058 (2017).


%\cite{Kocherlakota:2021dcv}
\bibitem{Kocherlakota:2021dcv}
P.~Kocherlakota \textit{et al.},
%``Constraints on black-hole charges with the 2017 EHT observations of M87*,''
Phys. Rev. D \textbf{103}, 104047 (2021).

%\cite{Bambi:2019tjh}
\bibitem{Bambi:2019tjh}
C.~Bambi, K.~Freese, S.~Vagnozzi and L.~Visinelli,
%``Testing the rotational nature of the supermassive object M87* from the circularity and size of its first image,''
Phys. Rev. D \textbf{100}, 044057 (2019).

%\cite{Johannsen:2010ru}
\bibitem{Johannsen:2010ru} 
T.~Johannsen and D.~Psaltis,
%``Testing the No-Hair Theorem with Observations in the Electromagnetic Spectrum: II. Black-Hole Images,''
Astrophys.\ J.\  {\bf 718}, 446 (2010).

%\cite{Johannsen:2015qca}
\bibitem{Johannsen:2015qca} 
T.~Johannsen,
%``Photon Rings around Kerr and Kerr-like Black Holes,''
Astrophys.\ J.\  {\bf 777}, 170 (2013).



\end{thebibliography}
\end{document}